\documentclass[prb,showpacs,preprintnumbers,preprint,amsmath,amssymb]{revtex4}
\usepackage{graphicx}
\usepackage{dcolumn}
\usepackage{bm}

\newcommand{\PrOsSb}{PrOs$_4$Sb$_{12}$}
\newcommand{\PrRuSb}{PrRu$_4$Sb$_{12}$}
\newcommand{\PrOsRuSb}{Pr(Os$_{1-x}$Ru$_x$)$_4$Sb$_{12}$}
\newcommand{\Rufive}{Pr(Os$_{0.95}$Ru$_{0.05}$)$_4$Sb$_{12}$}

\begin{document}

\title{Evolution of crystalline electric field effects, superconductivity, and
heavy fermion behavior in the specific heat of \PrOsRuSb}

\author{N. A. Frederick}

\author{T. A. Sayles}

\author{M. B. Maple}

\affiliation{%
Department of Physics and Institute for Pure and Applied Physical
Sciences, University of California at San Diego, La Jolla, CA
92093
}%

\date{\today}

\begin{abstract}

Specific heat $C(T)$ measurements were made on single crystals of
the superconducting filled skutterudite series \PrOsRuSb{} down to
$0.6$ K.  Crystalline electric field fits in the normal state
produced parameters which were in agreement with previous
measurements.  Bulk superconductivity was observed for all values
of the Ru concentration $x$ with transition temperatures
consistent with previous experiments, confirming a minimum in
$T_{c}$ at $x=0.6$. The $C(T)$ data below $T_{c}$ appear to be
more consistent with power law behavior for $x=0$ (\PrOsSb), and
with exponential behavior for $0.05 \leq x \leq 0.2$.  An enhanced
electronic specific heat coefficient $\gamma$ was observed for $x
\leq 0.4$, further supporting $x \simeq 0.6$ as a critical
concentration where the physical properties abruptly change.
Significant enhancement of $\Delta C/T_{c}$ above the weak
coupling value was only observed for $x=0$ and $x=0.05$.

\end{abstract}

\pacs{65.40.Ba, 71.27.+a, 74.25.Bt, 74.62.Dh}

\keywords{}

\maketitle

\section{Introduction}

The filled skutterudite compound \PrOsSb{} has proven to be an
intriguing and unusual material since its discovery as the first
Pr-based heavy fermion superconductor a few years ago.
\cite{Bauer02a,Maple02a}  Both the heavy fermion ($m^{*} \sim
50~m_{e}$) and the superconducting ($T_{c} = 1.85$ K) states
display very unusual properties.  The ground state of the
Pr$^{3+}$ ion in \PrOsSb{} that arises from the splitting of the
Pr$^{3+} J = 4$ multiplet in a crystalline electric field (CEF) is
nonmagnetic, and is either a $\Gamma_{1}$ singlet or a
$\Gamma_{3}$ doublet. The $\Gamma_{5}$ triplet first excited state
is $\sim 10$ K above the ground state, with the other excited
states following at $\sim 100$ K ($\Gamma_{4}$ triplet) and $\sim
300$ K ($\Gamma_{3}$ or $\Gamma_{1}$, respectively).  In the
superconducting state, \PrOsSb{} exhibits multiple transitions in
specific heat \cite{Maple02a,Maple03a} and magnetic penetration
depth,\cite{Broun04} and may also contain multiple superconducting
phases.\cite{Izawa03}  The nature of the superconducting energy
gap is also not clear: muon spin rotation \cite{MacLaughlin02}
($\mu$SR) and Sb-nuclear quadrupole resonance \cite{Kotegawa03}
(Sb-NQR) measurements indicate isotropic and strong-coupling
superconductivity, tunneling spectroscopy measurements support a
nearly fully gapped but unconventional superconducting order
parameter,\cite{Suderow03} and data from thermal conductivity in a
magnetic field\cite{Izawa03} and magnetic penetration
depth\cite{Chia03} are consistent with point nodes in the energy
gap. Additional $\mu$SR measurements reveal possible time-reversal
symmetry breaking in the superconducting state, further suggesting
that the superconducting state does not have s-wave
symmetry.\cite{Aoki03}

\PrRuSb{} is a much simpler compound than \PrOsSb.  It is also
superconducting ($T_{c} = 1.1$ K), but displays more conventional
properties.\cite{Takeda00}  From Sb-NQR measurements, the
superconductivity appears to be weak-coupling with an isotropic
energy gap.\cite{Yogi03}  Magnetic penetration depth measurements
yield moderate coupling and a fully gapped order
parameter.\cite{Chia04}  In addition, \PrRuSb{} is not a heavy
fermion compound; it has an electronic specific heat coefficient
$\gamma \sim 10$ times smaller than that of \PrOsSb. Features in
the physical properties of \PrRuSb{} could be described by a CEF
model with a $\Gamma_{1}$ ground state and a $\Gamma_{4}$ first
excited state separated by $\sim 70$ K.\cite{Takeda00,Abe02}

The \PrOsRuSb{} series of compounds was previously studied through
measurements of magnetic susceptibility $\chi(T)$ and electrical
resistivity $\rho(T)$.\cite{Frederick04}  Superconductivity was
found to persist for all values of the Ru concentration $x$, with
a minimum in the $T_{c} - x$ phase diagram at $x = 0.6$ where
$T_{c} = 0.75$ K.  This minimum may arise from a competition
between the heavy fermion superconductivity of \PrOsSb{} ($x = 0$)
and the BCS superconductivity of \PrRuSb{} ($x = 1$).  Based on
theoretical models, it has recently been suggested that there may
be a mixed-parity superconducting state near this minimum in
$T_{c}$.\cite{Sergienko04}  CEF effects were also observed for all
values of $x$ in the normal state of $\chi(T)$ and $\rho(T)$, with
the splitting between the ground state and the first excited state
increasing monotonically with $x$ between $x = 0$ and $x = 1$. For
$\chi(T)$, fits with a $\Gamma_{3}$ ground state were consistent
with the data for all values of $x$, while fits with a
$\Gamma_{1}$ ground state were only satisfactory near the extremal
values of the \PrOsRuSb{} series. The $\rho(T)$ data were also fit
with CEF equations, and although the fits were insensitive to the
degeneracy of the ground state, they were still able to provide
level splittings consistent with those derived from the $\chi(T)$
data. In the present study, the specific heat $C(T)$ of
\PrOsRuSb{} was measured down to $\sim 0.6$ K, to further
investigate the normal and superconducting state properties of
this extraordinary system.

\section{Experimental Details}

The single crystal specimens of \PrOsRuSb{} investigated in this
work were identical to those previously studied.\cite{Frederick04}
Specific heat $C$ was measured as a function of temperature
between $0.6$ and $50$ K in a $^{3}$He semiadiabatic calorimeter
by using a standard heat pulse technique.  The samples were
attached to a sapphire platform with Apiezon N grease.  The data
presented in this work were taken from experiments on collections
of single crystals with total masses between $11$ and $114$ mg.
X-ray measurements show no signs of multiple phases in the doped
materials; however, there was some sample dependence of the
superconducting transition in electrical resistivity, especially
on the doped materials.\cite{Frederick04}  Thus, the $C(T)$ data,
especially at the superconducting transitions, are expected to be
slightly broadened by the sample dependence of the crystals.

\section{Results and Discussion}

\begin{figure}[tbp]
\begin{center}
\includegraphics[angle=270,width=3.5in]{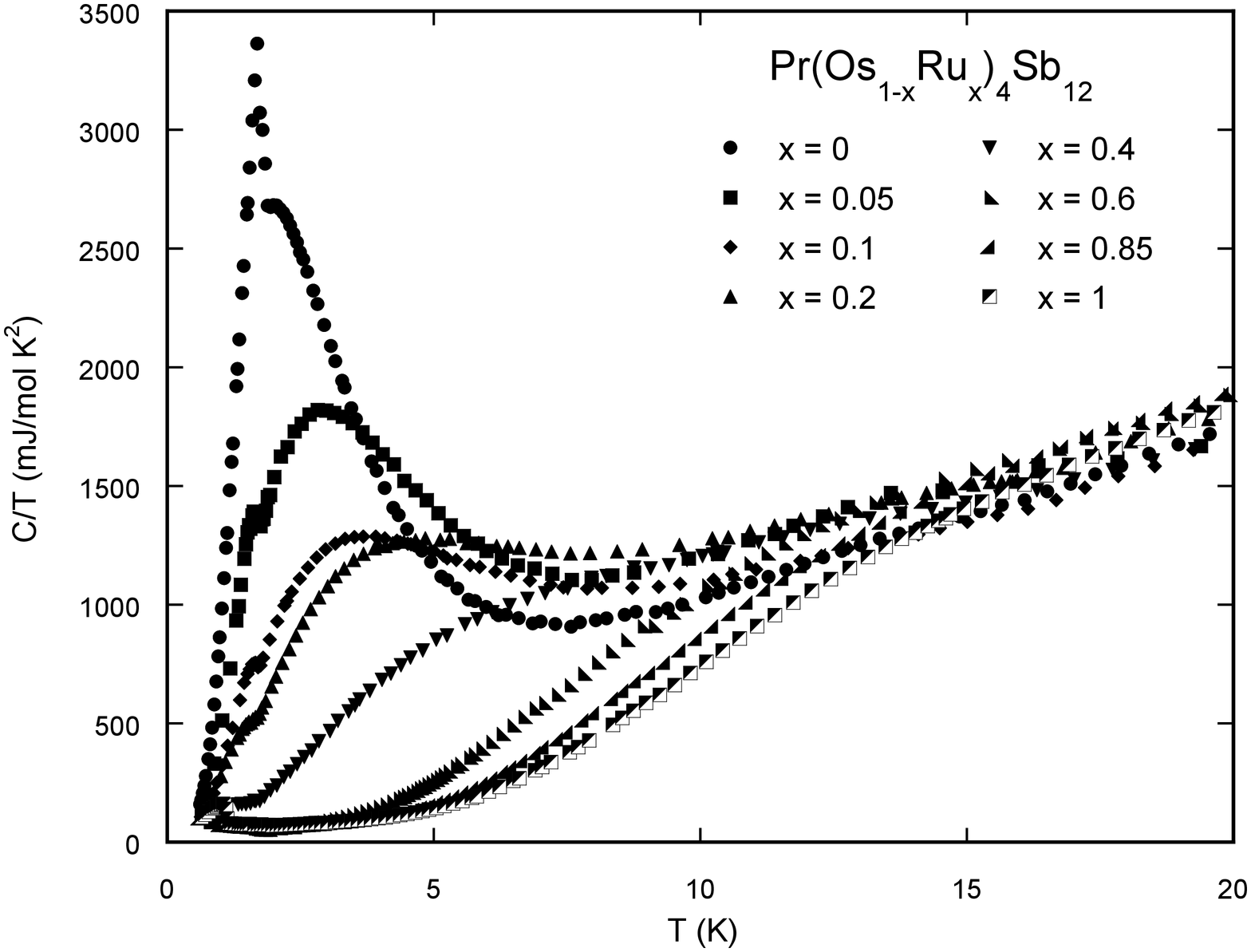}
\end{center}
\caption{Specific heat divided by temperature $C/T$ below $20$ K
for single crystal samples of \PrOsRuSb.} \label{CovT}
\end{figure}

Displayed in Fig.\ \ref{CovT} are specific heat divided by
temperature $C/T$ vs $T$ data for various Ru concentrations $x$ of
the \PrOsRuSb{} series for temperatures between $0.6$ and $20$ K.
The maximum of the Schottky anomaly, associated with the CEF
splitting of the Pr$^{3+}$ energy levels, noticeably decreases in
magnitude and shifts to higher temperatures with increasing $x$.
All of the $C(T)$ data for \PrOsRuSb{} were fitted between their
superconducting transitions $T_{c}$ and $10$ K by an equation
including electronic, lattice, and Schottky terms:
\begin{equation}
C/T = \gamma + \beta T^{2} + rC_{\rm Sch}/T. \label{Ctotal}
\end{equation}
Here $\gamma$ is the electronic specific heat coefficient, $\beta
\propto \Theta_{D}^{-3}$ is the lattice specific heat coefficient
(where $\Theta_{D}$ is the Debye temperature), and $C_{\rm
Sch}(T)$ is the Schottky specific heat anomaly for a two level
system arising from the energy difference between the CEF ground
state and the first excited state, scaled by a factor $r$. The
results of these fits are listed in Table \ref{CEFTable}.  These
fits find values of $\Theta_{D}$ for the end member compounds
comparable to other single crystal results of $165$ K for
\PrOsSb{}\cite{Vollmer03} and $232$ K for
\PrRuSb{}.\cite{Takeda00}

The Schottky specific heat anomaly $C_{\rm Sch}(T)$ for a
two-level system is given by
\begin{equation}
C_{\rm Sch}(T) = R\left(\frac{\delta}{T}\right)^{2}
\frac{g_{0}}{g_{1}}
\frac{\exp{(\delta/T)}}{[1+(g_{0}/g_{1})\exp{(\delta/T)}]^{2}},
\label{CSch}
\end{equation}
where $\delta$ is the energy difference in units of K between the
two levels, and $g_{0}$ and $g_{1}$ are the degeneracies of the
ground state and excited state, respectively.\cite{Gopal66} In
zero magnetic field, this equation is independent of whether or
not the local symmmetry of the Pr$^{3+}$ ions is cubic or
tetrahedral. It was found during the fitting procedure that in
order for the fits to be accurate, one or more of the terms in
Eq.\ \ref{Ctotal} had to be scaled. Fits were done both with $r$
modifying the Schottky term, and with $r$ multiplying the entire
equation.  The former case would be interpreted as some internal
broadening of the energy levels or as an overall transfer of
entropy to the itinerant electrons due to hybridization, while the
latter case would imply impurity phases (most likely free Sb)
causing an overall overestimate of the sample mass. While both
possibilities produced good qualitative fits, the values for
$\gamma$ resulting from assuming an overall scaling were extremely
large and not physically reasonable. Therefore, all the fits
presented here were exactly as shown in Eq.\ \ref{Ctotal}, with
$r$ only modifying the Schottky anomaly term.

The normal state fits were only performed up to $10$ K so that the
$C_{\rm lattice} \approx \beta T^{3}$ approximation would more
likely be accurate; however, the lattice terms are clearly the
smallest in this temperature range compared to the other terms,
and are thus difficult to accurately fit.  This appears to
especially be true for $x=0.2$ and $x=0.4$, where $\Theta_{D}$ is
suppressed compared to the end member compounds, and which may be
due to the disorder inherent in the substituted compounds. Because
of the uncertainty in the accuracy of the fit values for
$\Theta_{D}$, the error in the other parameters was estimated by
varying $\Theta_{D}$ by $\pm10$ and refitting the data.  The
largest error in the normal state arises in $\gamma$, while the
errors in $\delta$ and $r$ are much smaller.  These errors are
represented in Table \ref{CEFTable} and in the figures where
appropriate.

\begin{figure}[tbp]
\begin{center}
\includegraphics[width=3.5in]{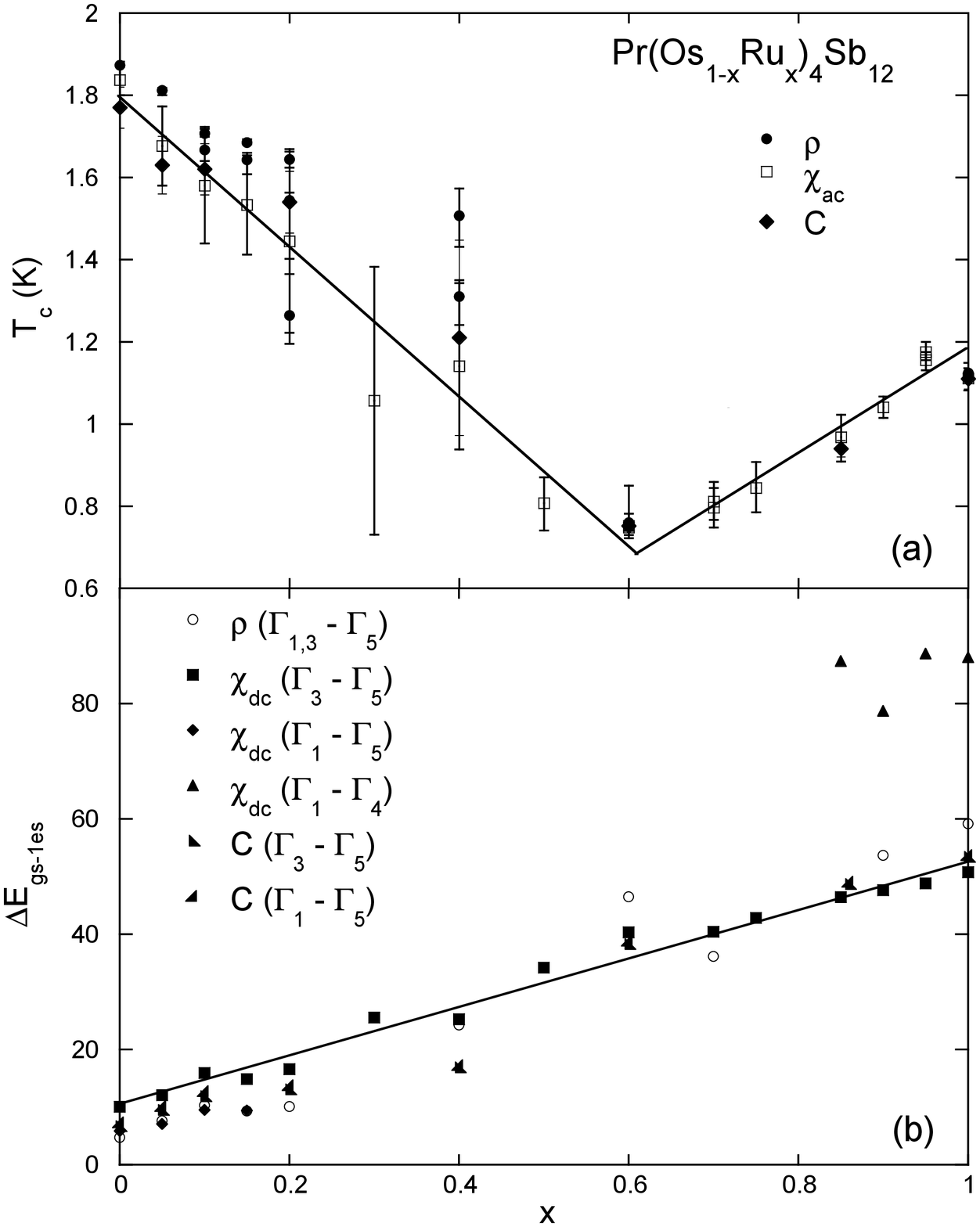}
\end{center}
\caption{(a) Superconducting critical temperature $T_{c}$ vs Ru
concentration $x$ for \PrOsRuSb, including data from measurements
of $\rho(T)$, $\chi_{\rm ac}(T)$, and $C(T)$. The straight lines
are guides to the eye.  (b) The splitting between the ground state
and first excited state $\Delta E_\mathrm{gs-1es}$ vs Ru
concentration $x$ for \PrOsRuSb, calculated from fits of CEF
equations to $\chi_\mathrm{dc}(T)$, $\rho(T)$, and $C(T)$ data.
The data points derived from $\rho(T)$, $\chi_{\rm ac}(T)$, and
$\chi_{\rm dc}(T)$ are from previous work.\cite{Frederick04}}
\label{Diagrams}
\end{figure}

The question of whether or not the ground state in \PrOsSb{} is a
$\Gamma_{3}$ doublet or a $\Gamma_{1}$ singlet has been
contentious since the heavy fermion superconductivity of \PrOsSb{}
was discovered.  In our original reports of heavy fermion
superconductivity in \PrOsSb,\cite{Bauer02a,Maple02a} fits to the
magnetic susceptibility yielded two possible Pr$^{3+}$ crystalline
electric field splittings, both with a $\Gamma_{5}$ first excited
state and either a $\Gamma_{1}$ singlet ground state or a
$\Gamma_{3}$ nonmagnetic doublet ground state with a quadrupole
moment.  At the present time, it appears that the overall data are
better explained by a $\Gamma_{1}$ singlet ground
state.\cite{Kohgi03,Rotundu04a,Goremychkin04} Nevertheless, it was
felt that fits to the $C(T)$ data for both possibilities should be
made. As can be seen from Table \ref{CEFTable} and Fig.\
\ref{Diagrams}(b), when fitting the normal state data up to $10$
K, both possible ground state fits result in reasonable values
that agree well with those previously published for \PrOsRuSb{}
based on $\chi(T)$ and $\rho(T)$ measurements.\cite{Frederick04}
However, for $x \leq 0.4$, the scaling factor $r$ is much closer
to $1$ for a $\Gamma_{3}$ ground state compared to $\Gamma_{1}$.
In fact, the best $\Gamma_{3}$ ground state fit for $x = 0$, pure
\PrOsSb, results in almost no scaling whatsoever. (The present
fits were performed on data from a measurement of single crystals,
while some previously published fits were performed on data from a
measurement of a pressed pellet.\cite{Bauer02a,Maple02a,Maple03b})
These results of $r_{\Gamma_{3}} = 1.01$ and $r_{\Gamma_{1}} =
0.56$ for $x=0$ are very consistent with fits performed on data
from measurements of single crystals by Vollmer {\it et al.},
which resulted in the values $r_{\Gamma_{3}} \approx 0.99$ and
$r_{\Gamma_{1}} \sim 0.5$.\cite{Vollmer03} As $x$ increases, the
scaling factor decreases, indicating a suppression of the Schottky
anomaly. At face value, the fact that $r_{\Gamma_{3}}$ is always
closer to $1$ than $r_{\Gamma_{1}}$ for $x \leq 0.4$ could be
considered as support for a $\Gamma_{3}$ ground state. It has been
suggested that the suppression of the Schottky anomaly, especially
for a $\Gamma_{1}$ ground state fit for \PrOsSb, could result from
an energy dispersion due to Pr-Pr interactions\cite{Aoki02} or
hybridization between the Pr f-electrons and ligand
states.\cite{Rotundu04b} As these arguments can apply equally well
to either ground state, these normal state results for $x \leq
0.4$ appear unable to discern between $\Gamma_{1}$ and
$\Gamma_{3}$ ground states.

For $x > 0.4$, the scaling factor $r$ increases rapidly. \PrRuSb{}
exhibits a situation complementary to that of \PrOsSb: while
$r_{\Gamma_{3}}$ is still $1.8-1.9$ times larger than
$r_{\Gamma_{1}}$, it is $r_{\Gamma_{1}}$ that is closer to $1$.
The accuracy of these results could be affected by the temperature
limits of the fits; for these large splittings, the maximum of the
Schottky anomaly is well above $10$ K.  However, the calculated
values of $\delta$ for the different ground states agree very well
overall with the values previously measured,\cite{Frederick04}
supporting the monotonic increase of the splitting between the
ground state and the first excited state throughout the
\PrOsRuSb{} series (Fig.\ \ref{Diagrams}(b)).

\begin{figure}[tbp]
\begin{center}
\includegraphics[width=3.5in]{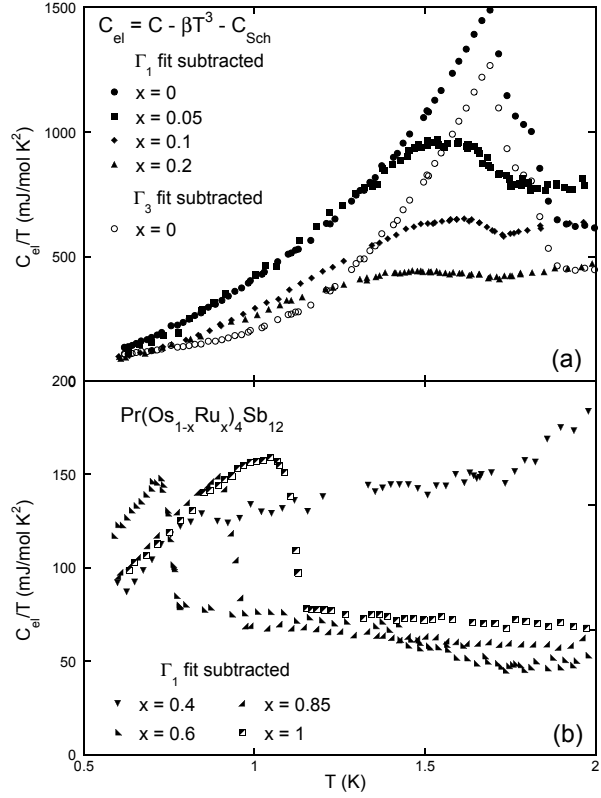}
\end{center}
\caption{The electronic specific heat divided by temperature
$C_{el}/T$ ($C_{\rm el} = C - C_{\rm lat} - C_{\rm Sch}$) of
\PrOsRuSb{} below $2$ K for (a) $0 \leq x \leq 0.2$ and (b) $0.4
\leq x \leq 1$. The data that were subtracted came mostly from
fits including a $\Gamma_{1}$ ground state.  The data from
subtracting a $\Gamma_{3}$ ground state fit were only included for
$x=0$ due to large differences that are not present for other
concentrations.} \label{CSC}
\end{figure}

In order to get a more accurate determination of $T_{c}$ and the
specific heat jump at $T_{c}$, $\Delta C/T_{c}$, the CEF and
lattice fit results were subtracted from the data at low
temperatures, leaving only the electronic specific heat.  The
subtractions were carried out using both the $\Gamma_{3}$ and the
$\Gamma_{1}$ ground state CEF fits, including those due to the
variation of $\Theta_{D}$ in order to further estimate the error
of the parameters. All the subtractions resulted in exactly the
same $T_{c}$ values, and very similar (within experimental error)
$\Delta C/T_{c}$ values, which are both listed in Table
\ref{SCTable}. These values of $T_{c}$ are plotted as a function
of $x$ in Fig.\ \ref{Diagrams}(a), along with previously measured
values; the error bars for the $T_{c}$ points represent the width
of the transitions in the respective measurements.  All the data
agree very well, and the minimum at $x=0.6$ is also reproduced in
the current data. With the exception of pure \PrOsSb{} (i.e., for
$x>0$), the shapes of the measured $C(T)$ curves below $T_{c}$
were nearly the same for either $\Gamma_{1}$ or $\Gamma_{3}$-based
subtraction. However, for \PrOsSb, a significant difference can be
seen in the slope of the data below $T_{c}$ for the two different
subtractions.  The $C(T)$ data after subtraction of $C_{\rm
lat}(T)$ and $C_{\rm Sch}(T)$ for a $\Gamma_{1}$ or $\Gamma_{3}$
ground state, $C_{\rm el}(T)$, are shown in Fig.\ \ref{CSC}. All
the concentrations are shown for the $\Gamma_{1}$ ground state
subtraction, and the $\Gamma_{3}$ ground state subtraction is also
shown for $x=0$.

\begin{figure}[tbp]
\begin{center}
\includegraphics[angle=270,width=3.5in]{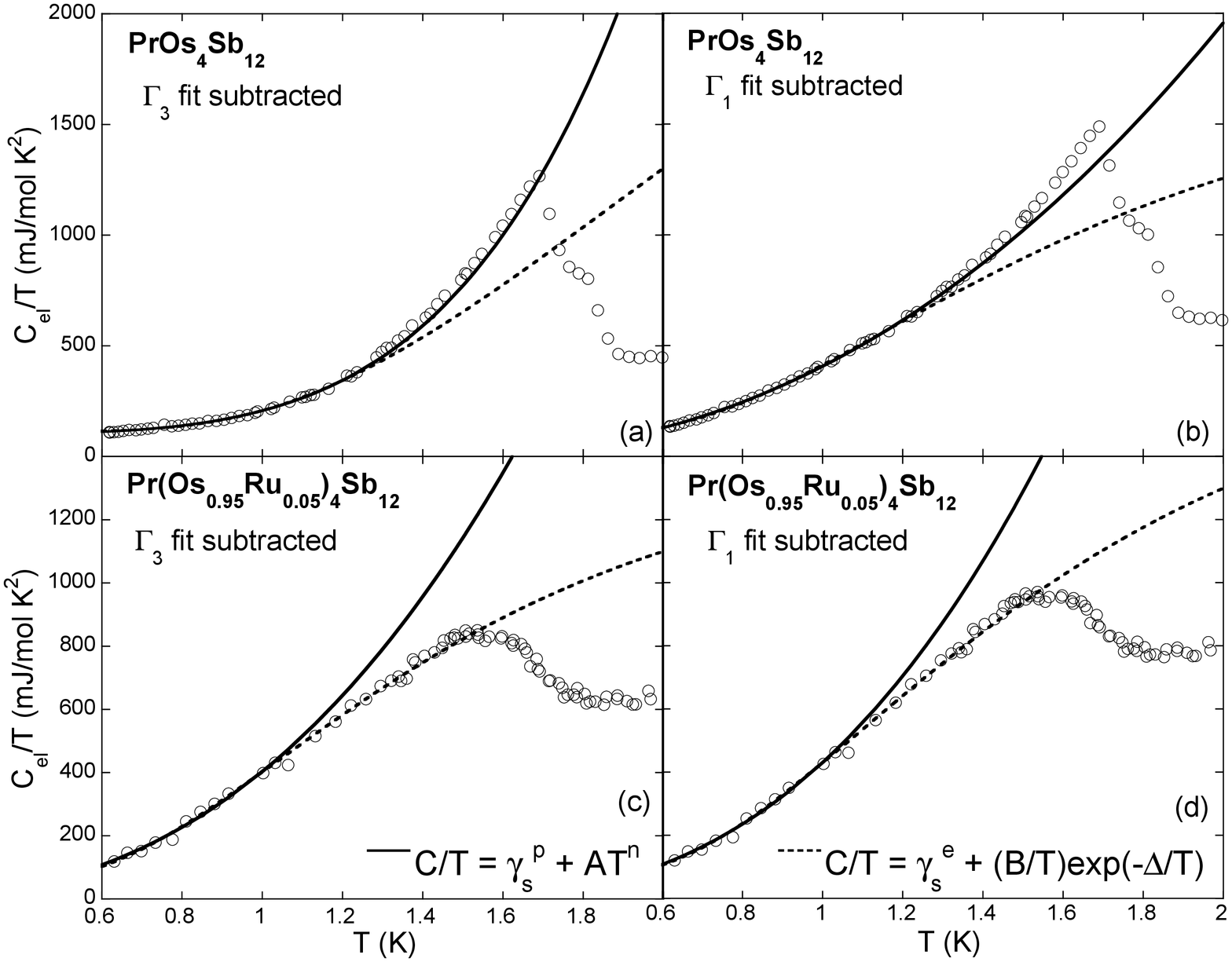}
\end{center}
\caption{Comparison of exponential (dashed line) and power law
(solid line) fits, subtracting data for either $\Gamma_{3}$ ((a)
and (c)) or $\Gamma_{1}$ ((b) and (d)) ground states below
$T_{c}$, for \PrOsSb{} ((a) and (b)) and \Rufive{} ((c) and (d)).
The \PrOsSb{} fits only extend up to $\sim 1.2$ K, and the
\Rufive{} fits only extend up to $\sim 1.1$ K, as described in the
text.} \label{SCcompare}
\end{figure}

The data below $T_{c}$, after the lattice and CEF terms were
subtracted, were fit to both power-law and exponential functions,
for energy gaps with and without nodes, respectively.  These
functions are typically considered to be valid only at very low
temperatures.  However, there are several examples of heavy
fermion superconductors which appear to display power-law behavior
up to near $T_{c}$ (e.g., \cite{Stewert84,Brison94}). The current
experiment had a low-temperature limit of $0.6$ K, which is
effectively a base temperature due to the large nuclear Schottky
contribution at lower temperatures.\cite{Vollmer03,Aoki02} The
high temperature limit of the fit was chosen to be
$\frac{2}{3}T_{c}$ in light of the above referenced examples, and
also to avoid possible spurious effects due to the width of the
superconducting transitions. Because of these constraints, only
the samples with $0 \leq x \leq 0.2$ could be fitted below
$T_{c}$, as there were not enough data points at the other
concentrations to give a reliable fit.  The power-law fit was of
the form
\begin{equation}
C/T = \gamma_{\rm s}^{\rm p} + AT^{n}, \label{Cpower}
\end{equation}
suggested for energy gaps containing nodes,\cite{Sigrist91} and
the exponential fit was of the BCS form\cite{Gopal66}
\begin{equation}
C/T = \gamma_{\rm s}^{\rm e} +
\frac{B}{T}\exp\left(-\frac{\Delta_{\rm e} }{T}\right),
\label{Cexp}
\end{equation}
where $\gamma_{\rm s}$ represents an electronic specific heat
coefficient in the superconducting state, $A$ and $B$ are fitting
constants, and $\Delta_{\rm e}$ is proportional to the energy gap
in the BCS theory of superconductivity.  The results of the
application of these fits to the $x=0$ and $x=0.05$ data, for both
ground state subtractions, are shown in Fig.\ \ref{SCcompare},
where the solid lines correspond to the power law fits and the
dashed lines represent the exponential fits. For $x=0$, the
power-law fits extrapolate from the highest fit temperature of
$\sim 1.2$ K all the way up to the transition, while the
exponential fits deviate from the data right above $1.2$ K. For $x
=0.05$ and higher, the converse is true, as the exponential fits
extrapolate to the higher temperature data (above $\sim 1.1$ K for
$x = 0.05$ and $x = 0.1$, and above $\sim 1.0$ K for $x = 0.2$)
much more accurately than the power law fits. While these
extrapolations cannot be taken by themselves as proof of the
superiority of one fit over the other, they are certainly
suggestive and intriguing. The broadened superconducting
transitions in the Ru-doped samples in particular may make the
exponential fit appear more appropriate.

\begin{figure}[tbp]
\begin{center}
\includegraphics[angle=270,width=3.5in]{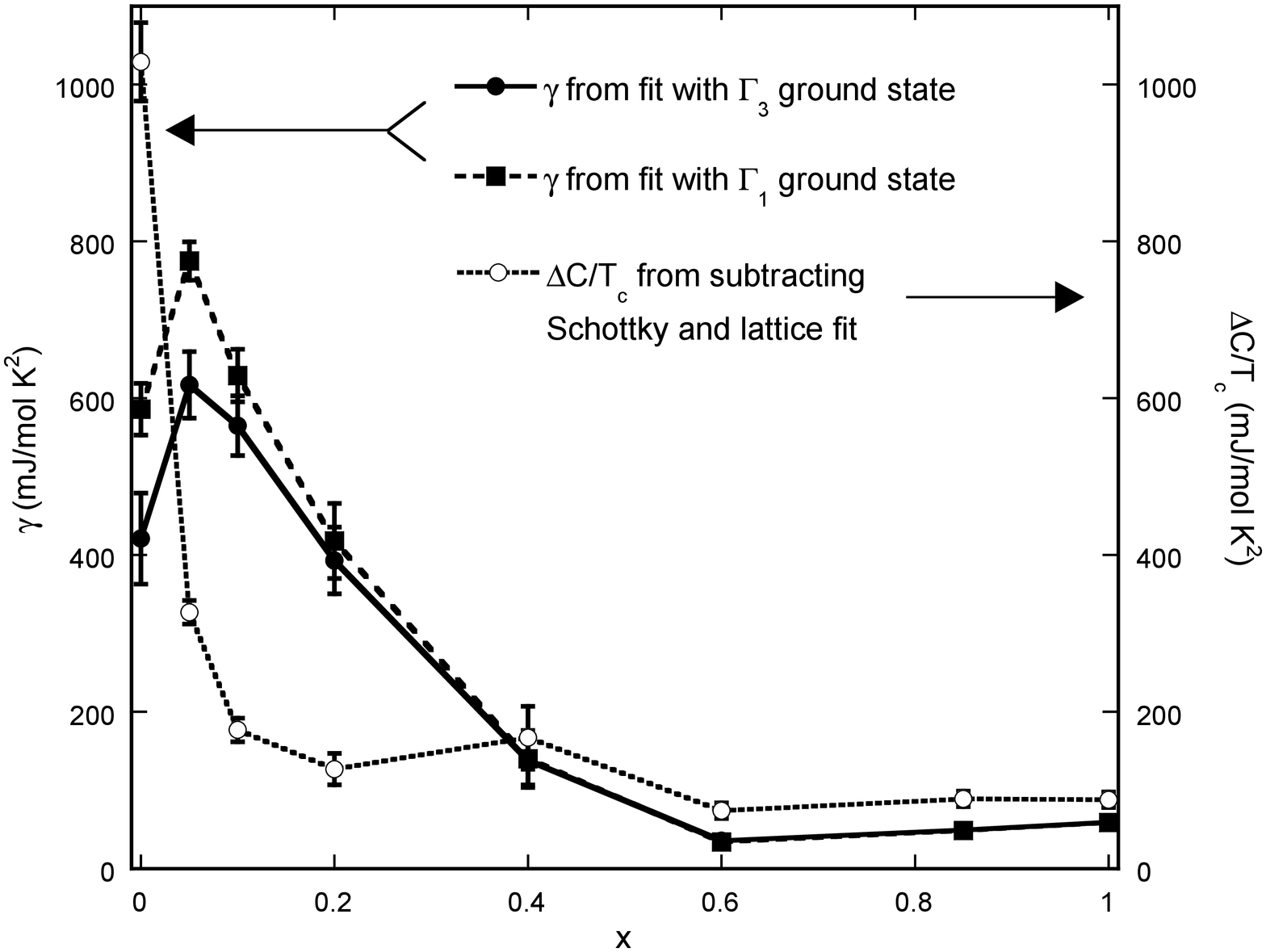}
\end{center}
\caption{Electronic specific heat coefficient $\gamma$ (left axis,
closed circles and squares) and $\Delta C/T_{c}$ (right axis, open
circles) as a function of Ru concentration $x$ for \PrOsRuSb.}
\label{gamma}
\end{figure}

The constants from the fits below $T_{c}$ are also listed in Table
\ref{SCTable}.  The listed errors are due to the variation of
$\Theta_{D}$ in the normal state fits.  The error in $T_{c}$ was
taken to be the width of the superconducting transition and is
presented graphically in Fig.\ \ref{Diagrams}a; the error in
$\Delta C/T_{c}$ from the variation of $\Theta_{D}$ was negligibly
small compared to the error inherent in the equal area
construction for determining $\Delta C/T_{c}$, which is
represented graphically in Fig.\ \ref{gamma}. Since these fits are
phenomenological, the absolute values of the resulting parameters
should not necessarily be trusted. However, comparing the fits and
the samples to one another can prove instructive. For pure
\PrOsSb, the $\Gamma_{3}$ ground state subtraction results in fits
that are much different from the $\Gamma_{1}$ ground state
subtraction results.  The other concentrations have similar fit
parameters for the two different ground states.  In addition, the
values for $\gamma_{\rm s}^{\rm p}$, $n$, and $\Delta_{\rm e}$,
and the errors associated with them, are much larger for the
$\Gamma_{3}~x=0$ fits than for the other $\Gamma_{3}$ fits,
falling well outside the spread of the other three data points. If
the fits are indeed accurate, then the value of $n$ for the
$\Gamma_{1}~x=0$ data is comparable with the $n=2$ expected in
$C/T$ for a heavy fermion compound with point nodes in the energy
gap.\cite{Sigrist91}  The values of $\Delta_{\rm e}$ for the
$\Gamma_{1}$ data of $\Delta_{\rm e}/T_{c} \approx 2.2-2.6$ are
moderately enhanced compared to the weak-coupling value of
$\Delta_{\rm e}/T_{c} = 1.76$ (for $\Delta_{\rm e}$ in units of
K). The parameter $\gamma_{\rm s}$ can be interpreted as a portion
of the sample that is normal or gapless; however, the values of
$\gamma_{\rm s}$ could also simply be artifacts of the fit,
especially in the case of the power law fit for $x=0.1$. The large
discrepancy between the fits below $T_{c}$ for \PrOsSb{}, even if
the absolute values are not entirely accurate, do lend
phenomenological support for $\Gamma_{1}$ being the ground state
in this compound. It is interesting that the analysis of $C(T)$ in
the superconducting state for samples containing a small amount of
Ru shows only small sensitivity to the choice of $\Gamma_{1}$ or
$\Gamma_{3}$ for the Pr$^{3+}$ ground state.

Fig.\ \ref{gamma} shows a comparison between the electronic
specific heat coefficient $\gamma$ calculated from the fits of the
normal state $C/T$ data and a value estimated from $\Delta
C/T_{c}$.  In the BCS theory of superconductivity, $\Delta
C/\gamma{}T_{c} = 1.43$.  This is not likely to be true in the
case of the unconventional superconductivity in \PrOsSb. However,
it is expected that there will be some proportional relation
between $\Delta C/T_{c}$ and $\gamma$ (e.g., $\Delta C/T_{c}
\propto \gamma$), and so it can still be instructive to view this
graphically. It can be seen in Table \ref{SCTable} and Fig.\
\ref{gamma} that $\gamma(x)$ derived from the normal state fits
exhibits a peak at $x=0.05$, decreases with $x$ to a minimum at
$x=0.6$, and then slowly rises with $x$ to $x=1$. In contrast,
$\Delta C/T_{c}$ starts out extremely large for $x=0$ and
decreases very quickly, again to a minimum value at $x=0.6$. The
``hump'' in the data at $x=0.4$ is due to the extremely broad
superconducting transition at this concentration.

The discrepancy between the values of $\gamma$ determined from the
normal state data and calculated from the superconducting
transition is an interesting one.  The minimum in $\gamma$ at
$x=0.6$ for the measured $\gamma$, along with the minimum in
$T_{c}$, strongly suggests that something unusual is happening
with the physical properties at this concentration.  In contrast,
the enhanced $\Delta C/T_{c}$ for $x=0$ and $x=0.05$ imply that
strong-coupling superconductivity is only present for these two
concentrations. Taken at face value, this could mean that for $0.1
\leq x \leq 0.4$, the heavy electrons are {\it not} participating
in the superconductivity, which would imply that the
superconductivity is nearly conventional for $x \geq 0.1$ (and
possibly also for $x=0.05$ if the exponential behavior below
$T_{c}$ is appropriate).  A more conventional explanation could be
that the Ru substitution broadens the superconducting transitions
in $C(T)$ enough to result in an underestimate for $\Delta
C/T_{c}$. This would only appear to be the case for $0.05 \leq x
\leq 0.4$, however, as the superconducting transitions for $x=0.6$
and $x=0.85$ are as sharp or sharper than for $x=1$.  It would be
unusual for disorder to play a strong role only for low Ru
concentrations.

In Figs.\ \ref{CSC} and \ref{SCcompare} it can be seen that the
sharp double superconducting transition present in pure \PrOsSb{}
is not obviously apparent in the Ru-doped samples, although this
may be obscured by the width of the transitions. Chia {\it et al.}
have recently suggested that this double transition may arise from
two superconducting phases which react differently to Ru
substitution; i.e., one is unconventional and is quickly destroyed
by impurities, while the other is more conventional and persists
throughout the entire series.\cite{Chia04}  It will be interesting
to see whether or not this conjecture can tie together the
seemingly contradictory results obtained through experiments over
the past few years. Indeed, in studies on the specific heat of
La-doped \PrOsSb, preliminary results suggest that only the lower
transition of \PrOsSb{} is suppressed for small La
concentrations.\cite{Rotundu04b}  This could also be the case for
the present Ru substitution studies.  On the other hand,
measurements of magnetic penetration depth suggest two-band
superconductivity in \PrOsSb, where the two bands are coupled by
Josephson pair tunnelling, which requires the two order parameters
to have the same symmetry.\cite{Broun04}  The difference in the
behavior below $T_{c}$ between \PrOsSb{} and the Ru-doped samples
could perhaps be accounted for by the suppression of one of the
superconducting bands.  Measurements on samples with small Ru
concentrations are underway in order to more closely track the
evolution of the superconducting properties in this remarkable
series of compounds.

\section{Summary}

The specific heat of single crystal samples of \PrOsRuSb{} was
measured down to $0.6$ K.  Fits to the normal state data resulted
in CEF parameters that monotonically increase with $x$, in
agreement with previously reported data. The superconducting
transitions in $C(T)$ were also consistent with previously
reported $\chi(T)$ and $\rho(T)$ data, confirming the minimum in
$T_{c}$ at $x=0.6$. Below $T_{c}$, a power law fit possibly
corresponding to point nodes in the energy gap could better
describe the data for \PrOsSb, but an exponential fit associated
with an isotropic energy gap was more appropriate for the data
with $0.05 \leq x \leq 0.2$. The electronic specific heat $\gamma$
was inferred from both normal state data and the ratio $\Delta
C/T_{c}$. The normal state fits revealed an enhanced $\gamma$ for
$x \leq 0.4$, reaching a minimum value at $x=0.6$, the same
concentration as the minimum in $T_{c}$. However, $\Delta C/T_{c}$
was only significantly enhanced for $x=0$ and $x=0.05$.

\section*{Acknowledgements}

We would like to thank T. D. Do, S. K. Kim, and D. T. Walker for
experimental assistance, and E. D. Bauer for useful discussions.
This research was supported by the U.S. Department of Energy Grant
No.~DE-FG02-04ER-46105, the U.S. National Science Foundation Grant
No.~DMR-03-35173, and the NEDO International Joint Research
Program.

\newpage

\begin{table}
\caption{Physical properties of samples of \PrOsRuSb, determined
from normal state specific heat data. The parameter $\delta$ is
the splitting between the ground state and the first excited state
in the Schottky anomaly, $r$ is the scaling factor for the
Schottky anomaly, $\gamma$ is the estimated electronic specific
heat coefficient, and $\Theta_{D}$ is the estimated Debye
temperature.  The errors in the parameters were determined by
allowing $\Theta_{D}$ to vary by $\pm10$ K within the fits (see
text for details).}\label{CEFTable}
\bigskip
\begin{tabular}{|c||cccc|cccc|}
\hline
 & \multicolumn{4}{c|}{$\Gamma_{3}$ ground state} &
 \multicolumn{4}{c|}{$\Gamma_{1}$ ground state} \\
$x$ & $\delta$ & $r$ & $\gamma$ & $\Theta_{D}$ & $\delta$ & $r$ &
$\gamma$ & $\Theta_{D}$ \\
 & (K) & & (mJ/mol K$^{2}$) & (K) & (K) & & (mJ/mol K$^{2}$) & (K) \\
\hline
0    & 6.72$\pm$0.06 & 1.01$\pm$0.03 & 421$\pm$58   & 186 & 7.36$\pm$0.04 & 0.56$\pm$0.01 & 586$\pm$33   & 211 \\
0.05 & 9.49$\pm$0.04 & 0.75$\pm$0.03 & 617$\pm$43   & 199 & 10.2$\pm$0.01 & 0.39$\pm$0.01 & 775$\pm$25   & 224 \\
0.1  & 11.9$\pm$0.2  & 0.55$\pm$0.02 & 565$\pm$38   & 200 & 12.8$\pm$0.2  & 0.29$\pm$0.01 & 629$\pm$34   & 203 \\
0.2  & 13.1$\pm$0.6  & 0.63$\pm$0.02 & 393$\pm$43   & 152 & 13.8$\pm$0.6  & 0.34$\pm$0.01 & 418$\pm$48   & 146 \\
0.4  & 16.9$\pm$1.4  & 0.45$\pm$0.09 & 138$\pm$31   & 139 & 17.3$\pm$1.6  & 0.23$\pm$0.05 & 140$\pm$36   & 135 \\
0.6  & 38.3$\pm$0.1  & 1.22$\pm$0.20 & 35.8$\pm$7.6 & 181 & 38.9$\pm$0.2  & 0.62$\pm$0.12 & 34.2$\pm$8.5 & 178 \\
0.85 & 48.7$\pm$0.6  & 2.23$\pm$0.09 & 49.3$\pm$4.7 & 218 & 49.1$\pm$0.6  & 1.15$\pm$0.05 & 48.7$\pm$4.9 & 216 \\
1.0  & 53.4$\pm$1.0  & 2.45$\pm$0.02 & 59.1$\pm$4.0 & 232 & 53.7$\pm$1.0  & 1.26$\pm$0.01 & 58.9$\pm$4.1 & 231 \\
\hline
\end{tabular}
\end{table}

\newpage

\begin{table}
\caption{Physical properties of samples of \PrOsRuSb, determined
from superconducting state specific heat data. The parameter $T_c$
is the superconducting transition temperature, $\Delta C$ is the
jump in $C(T)$ at $T_{c}$, $\gamma_{\rm s}$ is the electronic
specific heat coefficient in the superconducting state, $n$ is the
exponent of the power-law fit below $T_{c}$, and $\Delta_{\rm e}$
is the parameter in the exponential fit below $T_{c}$ that is
proportional to the energy gap. The errors in the parameters were
determined in the same manner as Table \ref{CEFTable}, by
propagating an error in $\Theta_{D}$, with the exception of
$T_{c}$ and $\Delta C/T_{c}$ (whose errors are represented
graphically; see text for details).}\label{SCTable}
\bigskip
\begin{tabular}{|c||c|ccccc|}
\hline
 & & \multicolumn{5}{c|}{$\Gamma_{3}$ ground state} \\
$x$ & $T_{c}$ & $\Delta C/T_{c}$ & $\gamma_{\rm s}^{\rm p}$ & $n$
& $\gamma_{\rm s}^{\rm e}$ & $\Delta_{\rm e}$ \\
 & (K) & (mJ/mol K$^2$) & (mJ/mol K$^2$) & &
 (mJ/mol K$^2$) & (K) \\
\hline
0    & 1.77 & 1021 & 101$\pm$1    & 4.55$\pm$0.40 & 114$\pm$1    & 6.46$\pm$0.43 \\
0.05 & 1.63 & 339  & 0            & 2.57$\pm$0.02 & 48.2$\pm$2.0 & 3.57$\pm$0.04 \\
0.1  & 1.62 & 179  & 23.0$\pm$1.9 & 2.59$\pm$0.04 & 73.2$\pm$1.0 & 4.15$\pm$0.03 \\
0.2  & 1.54 & 131  & 0            & 2.12$\pm$0.02 & 53.5$\pm$2.4 & 3.39$\pm$0.03 \\
0.4  & 1.21 & 165  &              &               &              &               \\
0.6  & 0.75 & 75   &              &               &              &               \\
0.85 & 0.94 & 89   &              &               &              &               \\
1.0  & 1.11 & 88   &              &               &              &               \\
\hline\hline
 & & \multicolumn{5}{c|}{$\Gamma_{1}$ ground state} \\
$x$ & $T_{c}$ & $\Delta C/T_{c}$ & $\gamma_{\rm s}^{\rm p}$ & $n$ &
$\gamma_{\rm s}^{\rm e}$ & $\Delta_{\rm e}$ \\
 & (K) & (mJ/mol K$^2$) & (mJ/mol K$^2$) & &
 (mJ/mol K$^2$) & (K) \\
\hline
0    & 1.77 & 1029 & 0            & 2.27$\pm$0.01 & 93.2$\pm$3.1 & 3.97$\pm$0.07 \\
0.05 & 1.63 & 327  & 0            & 2.71$\pm$0.01 & 55.2$\pm$0.7 & 3.79$\pm$0.01 \\
0.1  & 1.62 & 177  & 26.8$\pm$1.3 & 2.69$\pm$0.02 & 74.7$\pm$0.9 & 4.23$\pm$0.02 \\
0.2  & 1.54 & 127  & 0            & 2.13$\pm$0.02 & 52.9$\pm$2.7 & 3.42$\pm$0.02 \\
0.4  & 1.21 & 167  &              &               &              &               \\
0.6  & 0.75 & 74   &              &               &              &               \\
0.85 & 0.94 & 89   &              &               &              &               \\
1.0  & 1.11 & 88   &              &               &              &               \\
\hline
\end{tabular}
\end{table}

\newpage

\begin{thebibliography}{10}

\bibitem{Bauer02a}
E.~D. Bauer, N.~A. Frederick, P.-C. Ho, V.~S. Zapf, and M.~B.
Maple, Phys. Rev. B {\bf 65}, 100506(R) (2002).

\bibitem{Maple02a}
M.~B. Maple, P.-C. Ho, V.~S. Zapf, N.~A. Frederick, E.~D. Bauer,
W.~M. Yuhasz, F.~M. Woodward, and J.~W. Lynn, J. Phys. Soc. Jpn.
{\bf 71 Suppl.}, 23 (2002).

\bibitem{Maple03a}
M.~B. Maple, P.-C. Ho, N.~A. Frederick, V.~S. Zapf, W.~M. Yuhasz,
E.~D. Bauer, A.~D. Christianson, and A.~H. Lacerda, J. Phys.:
Condens. Matter {\bf 15}, S2071 (2003).

\bibitem{Broun04}
D.~M. Broun, P.~J. Turner, G.~K. Mullins, D.~E. Sheehy, X.~G.
Zheng, S.~K. Kim, N.~A. Frederick, M.~B. Maple, W.~N. Hardy, and
D.~A. Bonn, cond-mat/0310613.

\bibitem{Izawa03}
K. Izawa, Y. Nakajima, J. Goryo, Y. Matsuda, S. Osaki, H.
Sugawara, H. Sato, P. Thalmeier, and K. Maki, Phys. Rev. Lett.
{\bf 90}, 117001 (2003).

\bibitem{MacLaughlin02}
D.~E. MacLaughlin, J.~E. Sonier, R.~H. Heffner, O.~O. Bernal,
B.-L. Young, M.~S. Rose, G.~D. Morris, E.~D. Bauer, T.~D. Do, and
M.~B. Maple, Phys. Rev. Lett. {\bf 89}, 157001 (2002).

\bibitem{Kotegawa03}
H. Kotegawa, M. Yogi, Y. Imamura, Y. Kawasaki, G. q.~Zheng, Y.
Kitaoka, S. Ohsaki, H. Sugawara, Y. Aoki, and H. Sato, Phys. Rev.
Lett. {\bf 90}, 027001 (2003).

\bibitem{Suderow03}
H. Suderow, S. Vieira, J.~D. Strand, S. Bud'ko, and P.~C.
Canfield, Phys. Rev. B {\bf 69}, 060504(R) (2004).

\bibitem{Chia03}
E.~E.~M. Chia, M.~B. Salamon, H. Sugawara, and H. Sato, Phys. Rev.
Lett. {\bf 91}, 247003 (2003).

\bibitem{Aoki03}
Y. Aoki, A. Tsuchiya, T. Kanayama, S.~R. Saha, H. Sugawara, H.
Sato, W. Higemoto, A. Koda, K. Ohishi, K. Nishiyama, and R.
Kadono, Phys. Rev. Lett. {\bf 91}, 067003 (2003).

\bibitem{Takeda00}
N. Takeda and M. Ishikawa, J. Phys. Soc. Jpn. {\bf 69}, 868
(2000).

\bibitem{Yogi03}
M. Yogi, H. Kotegawa, Y. Imamura, G. q.~Zheng, Y. Kitaoka, H.
Sugawara, and H. Sato, Phys. Rev. B {\bf 67}, 180501(R) (2003).

\bibitem{Chia04}
E.~E.~M. Chia, M.~B. Salamon, H. Sugawara, and H. Sato, Phys. Rev.
B {\bf 69}, 180509(R) (2004).

\bibitem{Abe02}
K. Abe, H. Sato, T.~D. Matsuda, T. Namiki, H. Sugawara, and Y.
Aoki, J. Phys.: Condens. Matter {\bf 14}, 11757 (2002).

\bibitem{Frederick04}
N.~A. Frederick, T.~D. Do, P.-C. Ho, N.~P. Butch, V.~S. Zapf, and
M.~B. Maple, Phys. Rev. B {\bf 69}, 024523 (2004).

\bibitem{Sergienko04}
I.~A. Sergienko, Phys. Rev. B {\bf 69}, 174502 (2004).

\bibitem{Vollmer03}
R. Vollmer, A. Fai$\beta$t, C. Pfleiderer, H. v.~L\"{o}hneysen,
E.~D. Bauer, P.-C. Ho, V.~S. Zapf, and M.~B. Maple, Phys. Rev.
Lett. {\bf 90}, 057001 (2003).

\bibitem{Gopal66}
E.~S.~R. Gopal, {\em Specific Heat at Low Temperatures} (Plenum
Press, New York, 1966).

\bibitem{Kohgi03}
M. Kohgi, K. Iwasa, M. Nakajima, N. Metoki, S. Araki, N.
Bernhoeft, J.-M. Mignot, A. Gukasov, H. Sato, Y. Aoki, and H.
Sugawara, J. Phys. Soc. Jpn. {\bf 72}, 1002 (2003).

\bibitem{Rotundu04a}
C.~R. Rotundu, H. Tsujii, Y. Takano, B. Andraka, H. Sugawara, Y.
Aoki, and H. Sato, Phys. Rev. Lett. {\bf 92}, 037203 (2004).

\bibitem{Goremychkin04}
E.~A. Goremychkin, R. Osborn, E.~D. Bauer, M.~B. Maple, N.~A.
Frederick, W.~M. Yuhasz, F.~M. Woodward, and J.~W. Lynn, Phys.
Rev. Lett. {\bf 93}, 157003 (2004).

\bibitem{Maple03b}
M.~B. Maple, P.-C. Ho, N.~A. Frederick, V.~S. Zapf, W.~M. Yuhasz,
and E.~D. Bauer, Acta Physica Polonica B {\bf 34}, 919 (2003).

\bibitem{Aoki02}
Y. Aoki, T. Namiki, S. Ohsaki, S.~R. Saha, H. Sugawara, and H.
Sato, J. Phys. Soc. Jpn. {\bf 71}, 2098 (2002).

\bibitem{Rotundu04b}
C.~R. Rotundu, P. Kumar, and B. Andraka, cond-mat/0402599, (2004).

\bibitem{Stewert84}
G.~R. Stewert, Rev. Mod. Phys. {\bf 56}, 755 (1984).

\bibitem{Brison94}
J.~P. Brison, N. Keller, P. Lejay, A. Huxley, L. Schmidt, A.
Buzdin, N.~R. Bernhoeft, I. Mineev, A.~N. Stepanov, J. Flouquet,
D. Jaccard, S.~R. Julian, and G. G. Lonzarich, Physica B {\bf
199\&200}, 70 (1994).

\bibitem{Sigrist91}
M. Sigrist and K. Ueda, Rev. Mod. Phys. {\bf 63}, 239 (1991).

\end{thebibliography}
\end{document}